\documentclass[12pt,preprint]{aastex}

\slugcomment{ApJ, in press (submitted 5/11/07; accepted 7/17/07)}

\begin{document}

\title{Chemical Abundances of Luminous Cool Stars in the Galactic 
Center from High-Resolution Infrared Spectroscopy} 

\author{Katia Cunha\altaffilmark{1}}
\affil{National Optical Astronomy Observatory, P.O. Box 26732, Tucson, AZ
85726, USA; kcunha@noao.edu} 

\author{Kris Sellgren}
\affil{Department of Astronomy, The Ohio State University} %, 140 West 
%18th Avenue, Columbus, OH 43210,
%USA; sellgren@astronomy.ohio-state.edu}

\author{Verne V. Smith}
\affil{National Optical Astronomy Observatory} %, P.O. Box 26732, Tucson,
%AZ 85726, USA; vsmith@noao.edu}

\author{Solange V. Ramirez}
\affil{Infrared Processing and Analysis Center,
California Institute of Technology} %,
%Mail Code 100-22, Pasadena, CA 91125
%solange@ipac.caltech.edu}

\author{Robert D. Blum}
\affil{National Optical Astronomy Observatory} %, P.O. Box 26732, Tucson,
%AZ 85726, USA; rblum@noao.edu}

\author{Donald M. Terndrup}
\affil{Department of Astronomy, The Ohio State University} %, 140 West 
%18th Avenue, Columbus, OH 43210
%USA; terndrup@astronomy.ohio-state.edu}

\altaffiltext{1}{on leave from Observat\'orio Nacional, Rio de Janeiro, Brazil}

\begin{abstract}

We present chemical abundances in a sample of
luminous cool stars located within 30 pc of
the Galactic Center.  Abundances of carbon, nitrogen, oxygen,
calcium, and iron were derived from
high-resolution infrared spectra in the H- and K-bands.
The abundance results indicate that
both [O/Fe] and [Ca/Fe]
are enhanced respectively by averages of +0.2 and +0.3 dex,
relative to either the Sun or the Milky Way disk at near solar
Fe abundances.
The Galactic Center stars show a nearly uniform and nearly solar
iron abundance.
The mean value of A(Fe) = 7.59
$\pm$ 0.06 agrees well with previous
work. 
The total range in Fe abundance among Galactic Center stars,
0.16 dex,
is significantly narrower than the iron abundance distributions 
found in the literature for the older bulge population.  
Our snapshot of the current-day Fe abundance
within 30 pc of the Galactic Center
samples stars with an age less than 1 Gyr;
a larger sample in time (or space)
may find a wider spread in abundances.

\end{abstract}


%\clearpage

\section{Introduction}

Stellar abundances can provide critical information in understanding the 
formation and chemical
evolution of the halo, disk, and bulge. With the advent of 
infrared imaging and spectroscopy on large
telescopes, it is now possible to explore the stellar population 
residing near
the Galactic Center (within $\sim$200 pc) in detail.
Infrared techniques are essential because this region is obscured
by at least A$_{\rm V}$=30 mag. The Galactic
Center contains many luminous, massive stars. 
Those known to date are concentrated in 
three separate clusters within the central 60 pc,
the Central Cluster, the Arches Cluster, and the Quintuplet
Cluster, 
at (projected) galactocentric radii $R_G$ of
0, 24, and 30 pc, respectively.
We adopt a distance to the Galactic Center of 8.0 kpc (Reid 1993).
The youngest stars in these three clusters have ages of 3 -- 9 Myr 
(Najarro et al. 1994; Krabbe et al.
1995; Blum et al. 1996b), 1 -- 4.5 Myr, and 3 -- 5 Myr 
(Figer et al. 1998, 1999a, 1999b; Blum et al. 2001),
respectively.

The Central Cluster, which defines the dynamical center of the nucleus,
has a central stellar density of 4x10$^{6}$ M$_{\odot}$ pc$^{-3}$ at a radius of $\sim$
0.3 pc (Ghez et al. 1998, Genzel et al. 2000).
The center is believed to contain a massive black hole
of 3--4 $\times$ 10$^{6}$ M$_{\odot}$ (Ghez et al. 2005;
Sch\"odel et al 2003). Blum et al. (1996a)
have determined J, H, K, and L photometry for the 1100
brightest stars in the Central Cluster (within 5 pc).
Using the observed J - H, H - K, and 
K - L colors and assumed intrinsic colors, they
determined the interstellar extinction at 2.2 $\mu$m, finding a 
mean value of 3.3 mag, similar to previous
results. The de-reddened K-band luminosity function 
of the Central Cluster (Blum et al. 1996a) points
to a striking excess of luminous stars compared to the bulge stellar 
population in Baade's window.
These luminous, young (a few Myr old) stars co-exist in the Central Cluster with 
older stars (Blum et al. 1996b, 2003; Blommaert et al. 2003). 
Thus the Galactic Center stellar population differs markedly from that in the bulge of the 
Milky Way, which is mostly made up of
stars with ages $\sim$10 Gyr, with few stars younger 
than 5 Gyr (Kuijken \& Rich 2002; Zoccali et al. 2003).

Earlier abundance analysis of the Galactic Center 
include the pioneering studies
by Carr et al. (2000) and  Ramirez et al. (2000) 
who analyzed high resolution (R=40,000) infrared spectra.
Carr et al. (2000) derived abundances for the Galactic Center
supergiant IRS 7, finding nearly solar iron abundance and CNO
patterns characteristic of dredge-up of CNO cycle products.
Ramirez et al. (2000) analyzed 10 stars in the Galactic 
Center and found the mean [Fe/H] 
to be nearly solar, with [Fe/H] = +0.12$\pm$ 0.22.  
Furthermore, the mean [Fe/H] of the Galactic
Center stars was very similar to the mean [Fe/H] of eleven 
cool, luminous stars in the 
solar neighborhood,
chosen to have temperatures and luminosities
similar to the Galactic Center stars.
Ramirez et al. (2000) also found
that the distribution of [Fe/H] in Galactic Center
stars is significantly narrower than the distribution
of [Fe/H] in the bulge,
emphasizing the differences between the stellar populations of the
bulge and center.

In this study we analyze the same sample stars 
as in the Ramirez et al. (2000) study, 
but with additional sets of high-resolution IR-spectra. 
All but one target star are members
of the Central Cluster and are located at distances of 
$R_{g}$ = 0.21 -- 2.2 pc (Blum et al. 1996$b$). In order to
sample a different location in the Galactic Center, one target (VR 5-7) 
is a member of the Quintuplet Cluster
at $R_{g}$ = 31 pc (Moneti et al. 1994). The goal is to include 
abundances from $\alpha$-capture elements, in order 
to begin to probe the abundance 
distribution patterns in Galactic Center stars.  The
studied elements include carbon, nitrogen, oxygen, and calcium.
In order to have a homogeneous and self-consistent 
analysis of abundance patterns,
we also re-derived Fe abundances for the sample stars.

\section{Observations and Data Reduction}

Observations were made at the 8--m Gemini--South telescope,
on Cerro Pach\'{o}n in Northern Chile.
Spectra were obtained with the NOAO infrared echelle
spectrograph Phoenix (Hinkle et al. 2003) as part of observing programs
GS--2002A--DD--1 and GS--2003A-DD--1. This spectrograph
has a resolution $\lambda / \Delta \lambda$ $\approx$ $R$ = 50,000
when used, as in these observations, with a slit width of 4
pixels (0.34$''$). Spectra were obtained on the nights of 2002 May 10,
2002 May 11, 2002 June 11, 2002 June 13, and 2003 July 29. Typical image
quality estimated on acquisition frames (1.5 or 2.3 $\mu$m) was
approximately 0.5$''$ to 1.0$''$, and some data were taken through thin
clouds.

Phoenix spectra were obtained of five Galactic Center stars (BSD72,
BSD114, BSD124, IRS11, IRS19)
selected from the sample of Ramirez et al. (2000), and of
one solar neighborhood S giant (HD172804), selected from the sample of
Smith \& Lambert (1985, 1986, 1990). Each star was observed at two
grating settings with central wavelengths of 1.5649 $\mu$m (filter
H6420) and 2.3153 $\mu$m (filter K4308). 
Stars of spectral types
A or B were also observed at an airmass similar to that of each star.
The spectra of these hot stars, which are featureless at the
wavelengths of interest, were later divided into the spectra
of the Galactic Center stars to remove telluric absorption features.

Each star was nodded along the slit, and spectra obtained at
two slit positions, for background subtraction. In the very
crowded stellar field of the Galactic Center, the nod spacing
was carefully chosen for each star to avoid another star
falling into the background position. In several cases, the source was
nodded off the slit in order to find a clear sky position. Flat fields were
obtained by observing the facility--provided flat field lamp (GCAL;
Phoenix uses a non--standard M3 position in the Gemini instrument
support structure).

All spectra were flat-fielded and background-subtracted, and
corrected for bad pixels. Each Galactic Center star spectrum was divided
by the appropriate hot star, with the same nod spacing along the
slit, to correct for telluric absorption. This also provided
some correction for fringing in the spectrograph. Wavelength
calibration was done using telluric lines. The two nod positions
for each star were reduced separately, rather than coadded, for
as long as possible through the analysis, as a control on
systematic uncertainties such as fringing. No flux calibration
was done.

Additional observations of ten Galactic Center stars from the
sample of Ramirez et al. (2000), and four solar neighborhood
stars from the sample of Smith \& Lambert (1985, 1986, 1990)
were made at the NASA Infrared Telescope Facility 3-m telescope
at Mauna Kea Observatory. Spectra were obtained with the
facility infrared echelle spectrograph, CSHELL, at $R$ = 40,000
(Tokunaga et al. 1990). These observations have been previously
described by Ramirez et al. (2000); in brief, the observing
technique was identical to that described above.
Figure 1 shows one sample spectrum obtained with for the 
Galactic Center star BSD72.

\section{Analysis}

\subsection{Effective Temperatures and Surface Gravities}

Ten Galactic Center red stars are analyzed here, with all
program stars included in the abundance study by Ramirez et al. (2000).
Effective temperatures were derived for these stars by Blum
et al. (2003) using spectral indices defined by both CO and H$_{2}$O
absorption. The use of spectral indices is necessitated by the extremely
large reddening towards the Galactic Center ($A_K$ $\sim$3),
which renders even infrared colors, such as (J-K), too
uncertain. The absorption indices defined by Blum et al. (2003) are measured relative
to spectral regions adjacent to the absorption itself and are
thus independent of reddening.

The CO index is defined largely by the
2-0 $^{12}$CO bandhead at a wavelength of 2.2935$\mu$m and is
sensitive to both T$_{\rm eff}$ and luminosity class (i.e., gravity),
with the CO absorption increasing with luminosity for a given
temperature. The H$_{2}$O index is composed of a broad
absorption feature stretching from the red end of the $H-$band into the blue part
of the $K-$band. Blum et al. (2003) measure the H$_2$O index
from $R$ = 600 spectra from 1.2 to 2.4 $\mu$m, and define
continuum points for their H$_2$O index at both the blue end
and the red end of the broad H$_2$O absorption band to mitigate any 
sensitivity of the H$_2$O index to reddening. The H$_2$O absorption in both M giants
and supergiants spans 
about the same range of values, while Mira variables exhibit 
considerably stronger absorption by H$_{2}$O. 
H$_{2}$O is expected to exhibit the opposite change with respect
to gravity compared to CO; the feature becomes weaker in stars of the
same spectral type but lower gravity (Kleinmann \& Hall 1986).
Long period variables (LPVs, e.g. Mira-type stars) exhibit considerably 
stronger absorption than normal giants or supergiants owing to the 
very cool temperatures of LPVs.

A comparison of CO absorption versus H$_{2}$O absorption
is shown in the top panel of Figure 2 for samples of standard
giants, supergiants, and Miras (taken from the values in
Blum et al. 2003). This type of plot segregates these three
stellar groups. Also shown in Figure 2 (top panel) are the
Galactic Center stars; these stars largely inhabit the
region that both overlap and is adjacent 
to the giants and supergiants, with 
the Galactic Center stars tending to have CO strengths more similar
to supergiants but with stronger H$_{2}$O.
Most of the lower luminosity giants tend to have both lower
overall CO and H$_{2}$O absorption. A simplified view of the
Galactic Center stars studied here would most closely associate
them with luminous giants or supergiants and not much overlap
with the Miras. The CO and H$_{2}$O absorption indices can be
used as a way to associate a spectral type with a particular
Galactic Center star: recall that the extreme and variable
reddening into the Galactic center precludes the use of even
infra-red colors to ascertain reliable effective temperatures.

Blum et al. (2003) relied primarily on a calibration of the CO index
with T$_{\rm eff}$ defined by M giants and supergiants, which define
two sequences in a T$_{\rm eff}$-CO plane that are separated by
about 300K at a given value of CO absorption (with the supergiants
being hotter). In this study, a somewhat different calibration was
tried and compared to that from Blum et al. (2003). There is a slight
dependence in both the CO and H$_{2}$O indices with the stellar C/O
ratios, when the measured C/O ratios are taken from Smith \& Lambert
(1990). These dependences are caused by the dominant role played by
the tightly bound CO molecule on the partial pressures of, not only
CO itself, but also H$_{2}$O. The stars analyzed are all of spectral
type M, where C/O $\le$ 1.0 and, for a given O-abundance, the CO
absorption increases with increasing C/O ratio. At the same time
as C/O increases, the amount of free oxygen decreases and the
subsequent absorption by H$_{2}$O decreases. Thus, there is a slight
positive slope of CO absorption and a slight negative slope of
H$_{2}$O as C/O increases. 

The somewhat modified method to estimate spectral types explored here
involves adding the CO and H$_{2}$O indices as a way of cancelling,
to some degree, C/O trends. The resulting relation 
between the sum of the CO and H$_{2}$O indices vs. spectral type, 
for the standard
supergiants in Blum et al. (2003), is illustrated in the bottom
panel of Figure 2. There is a well-defined relation for spectral
types later than M0, which will include all of the Galactic center
stars studied here. A straight line was fit to the absorption
index versus spectral type for M0 to M5 and this line is also
shown in the figure.

The spectral types derived from the H$_{2}$O + CO indices were then
mapped into effective temperatures 
via a T$_{\rm eff}$-spectral type relation
as shown in Figure 3. The supergiant points on this plot were
taken from the recent analysis of the red supergiant T$_{\rm eff}$-scale
by Levesque et al. (2005). The relation between spectral type
and effective temperature is quite well defined. Included in Figure 3
is the relation for M-giants of Smith \& Lambert (1985; 1986; 1990)
coupled to the accurate spectral types from Yamashita (1971).
These relations have involved no color information, but rely mainly
on the TiO bands in the visible and seem to present very similar
relations for both giants and supergiants. The two sequences appear
to begin to separate at spectral types later than about M4,
but almost no supergiants are found later than about M5 anyway.
The resulting relation
for the supergiants was used to estimate the spectral
types of the Galactic Center stars, based upon their measured
CO and H$_{2}$O indices; this relation was selected based
on the luminosities of the Galactic Center stars, as discussed above. 

The final T$_{\rm eff}$ adopted was an average of that derived here and
the temperature from Blum et al. (2003). The comparison between the
two estimates is good, with the mean and standard deviation being
$\Delta$(this study - Blum et al.)= +72$\pm$138K.
The scatter of $\sim$ $\pm$140K found between the two T$_{\rm eff}$-scales
can be compared to the expected inherent uncertainties in the 
T$_{\rm eff}$-scale adopted here based on errors in the fundamental
quantities. The error in the scale of H$_{2}$O + CO is about $\pm$5
which translates to an uncertainty in spectral type of $\pm$1.5 subtypes
(Figure 2). This uncertainty in spectral type leads to an expected error
of $\sim$ $\pm$150K from the relation plotted in Figure 3.
The effective temperatures adopted for the studied stars
are presented in Table 1 (Table 2 refers to the 
comparison disk stars described
below).

Surface gravities (noted as log g) for the Galactic Center stars are
also listed in Table 1. These values were derived by combining the
effective temperatures with the absolute bolometric magnitudes taken
from Blum et al. (2003). With the stars plotted in a M$_{\rm bol}$
versus T$_{\rm eff}$ diagram, stellar evolutionary model tracks are
added and each program stellar mass is estimated (column 7 of Table 1). 
The stellar model tracks are taken from the model 
grids from Salasnich et al.
(2000; which can be downloaded from 
http://pleiadi.pd.astro.it) and slightly ``supersolar'' metallicities
are used (Z=0.019, or, [m/H]=+0.1), which correspond fairly closely to
the final metallicities derived for the Galactic Center stars.
The M$_{bol}$-T$_{eff}$ diagram is illustrated in Figure 4. The Galactic
center stars are shown as the filled circles.
Representative stellar model evolutionary
tracks are also shown with the model mass labeled for each track.

As a test of the analysis techniques used for the Galactic Center
stars, four bright, ``local''disk M-giants and one S-giant were 
analyzed in the
same way using the same wavelength regions.  The five stars were HR6146,
HR6702, HR7442, HR8062 and HD172804.  
Their spectral types from Yamashita (1971)
were used with the spectral type-T$_{\rm eff}$ relation (Figure 3)
to determine their effective temperatures.  All five of these stars
have measured Hipparcos parallaxes, which were used to determine
distances and absolute V and K magnitudes.  Given their (V-K) colors,
the K-band bolometric corrections from 
Bessel et al. (1998) were applied to determine
M$_{\rm bol}$ for each M-giant or S-star.  In this way, surface gravities
were derived in the same way as for the Galactic Center stars in an
M$_{\rm bol}$-T$_{\rm eff}$ diagram and the estimated masses from the
Padua evolutionary tracks; these stars are represented as open squares
in Figure 4.  The derived stellar parameters for the
bright disk M-giants are listed in Table 2.

The ``nearby'' giants cluster around M$_{\rm bol}$ $\sim$ -4.1 to -4.7
and this roughly overlaps the 
lower-luminosity members of the Galactic Center
sample.  
The effective temperatures also overlap between the two samples.
Thus no large systematic differences are expected in the
derived abundances from the two samples.  
The bright M-giants constitute
a valuable control sample in interpreting the abundance patterns in the
Galactic Center stars.

Yusef-Zadeh \& Morris (1991)
and Serabyn et al. (1991) discovered that the outer envelope of
IRS 7 is ionized, and that it has a tail of ionized gas.
This has been interpreted as due to the interaction of the winds
from hot stars in IRS 16 and the mass-loss wind from IRS 7 (Serabyn et al.
1991; Yusef-Zadeh \& Melia 1992).
There is no evidence from
our high-resolution infrared spectra that the presence of a tail
of ionized gas affects in any measurable way the formation of the
photospheric spectra in IRS 7 (see also Carr et al. 2000).
Other wind-wind interactions have been discovered in the Central Cluster,
appearing as bow-shocks around suspected hot stars in the mid-infrared
(Tanner et al. 2005), but none of the other stars in our
Galactic Center sample are among these sources.
The photospheric absorption lines observed at high spectral
resolution from CO, OH, CN, Fe I, and Ca I, for the stars
in our Galactic Center sample, look normal when compared
to the Galactic disk control sample. In addition,
all of our Galactic Center sample stars have $H-$band and $K-$band
spectra at $R$ = 600--1200 (Blum et al. 1996, 2003), and some have
$K-$band spectra at $R$ =2500 (Sellgren et al. 1987). The only
differences from normal M supergiant/AGB star spectra that have been
noted can be attributed to abundance patterns that we infer later
from our high-resolution spectroscopy.
%As noted first for
%IRS7 (Yusef-Zadeh \& Morris 1991; Serabyn et al. 1991) and more recently in a larger
%sample of Galactic Center stars (Tanner et al. 2005), many of these stars
%exhibit bow shocks and 'cometary tails'. Tanner et al. (2005) interpret these
%bow shocks and tails as due to stellar winds interacting with ambient
%material and not due to interaction with an enhanced UV radiation environment
%at the Galactic Center. The photospheric absorption lines observed here
%at high spectral resolution from CO, OH, CN, FeI and Ca I look normal
%when compared to the Galactic disk control sample. There is no evidence
%from the high-resolution infrared spectra that the bow shocks affect
%in any measurable way the formation of the photospheric spectra
%in the Galactic Center stars. In addition, all of the Galactic Center
%stars studied here were observed previously by Sellgren et al. (1987)
%at lower spectral resolution (R=2500) near 2.2 $\mu$m and noted to have
%essentially normal M-supergiant/giant spectra.

\subsection{ Microturbulent Velocities}

The microturbulent velocity parameter ($\xi$) for the program stars 
was obtained using a
two-step process.  
First, the measured equivalent widths of the Fe I lines 
were used to derive an initial estimate of the microturbulence 
from the requirement that the Fe I 
equivalent widths be independent of the derived Fe abundances.  
This method would be sufficient if the lines were unblended.
The Galactic Center stars under analysis, however, 
have fairly crowded spectra and
one must check whether
contributions from CN lines and
other blending lines influence the Fe I equivalent widths.
As a second step, therefore,
the line list in Section 3.3.2 was used to compute 
a grid of synthetic spectra 
for a range of microturbulent velocities bracketing
the initial microturbulence value defined in the first step.
The observed and synthetic spectra were compared to find
the final microturbulence which best fit each star.
The values presented in Table 1 and 2 are the results 
from such a fine-tuning of $\xi$.

A direct comparison of our microturbulent velocities with 
the values derived by
Ramirez et al. (2000) shows good agreement: 
$\Delta \xi$ (this study - Ramirez et al.) = 
-0.0$\pm$0.3 km s$^{-1}$. In addition, 
Carr et al. (2000) obtained 
$\xi$=3.0 km s$^{-1}$ for IRS7, which is within the errors of our
adopted value (3.2 $\pm$ 0.3 km s$^{-1}$) for this star.
In Figure 5 the microturbulent velocities and surface gravities are 
compared with 
results from several different studies in the literature. 
Although there is significant scatter of the points 
in the figure, the general trend can be described as a 
rough correlation between the microturbulence parameter and the 
surface gravity of the star 
(as discussed in e.g., Ramirez et al. 2000). 
In particular, we note that
the microturbulent velocities in this study 
(represented by black pentagons) generally fall 
within the range of values defined for those stars 
with similar surface gravities;
and the well-studied supergiant $\alpha$ Ori 
(represented by the green open square)
falls within the scatter of points defined 
by the Galactic Center stars
which have supergiant-like luminosities.  

\subsection{Abundances}

The abundance analysis in this study was carried out 
via comparisons of synthetic 
spectra with the observed ones.  The model atmospheres adopted for the 
studied late type giants and supergiants are from the grid of 
spherical models by Plez (Plez 1992; Plez, Brett, \& Nordlund 1992).
The program MOOG (Sneden 1973) was used in order 
to compute LTE synthetic spectra 
given the required input model atmosphere defined by the parameters of
effective temperature (T$_{\rm eff}$), surface gravity (log g), and 
microturbulent velocity ($\xi$). The synthetic spectra were convolved with 
both a Gaussian function, representing the instrumental profile, 
as well as
a macroturbulent velocity function (Gray 1992) to properly match the 
observed profiles of the spectral lines via minimizing the residuals between the
observed and synthetic profiles. The macroturbulent velocities adopted
in the calculations are presented in column 5 of Table 1 and 2.

\subsubsection{Carbon, Nitrogen and Oxygen}

The C, N, and O abundances in this study were derived 
from observations of OH and CN
molecular lines present in the H-band region, 
as well as CO lines present in
the K-band region of the Phoenix spectra 
obtained for three Galactic Center stars:
BSD72, BSD124 and IRS19.  
In order to obtain a consistent solution for the molecular 
equilibrium involving CN, CO and OH molecules, 
we first synthesized the OH molecular lines in the H-band region.  
The OH 
lines analyzed here are listed in Table 3 and, 
except for the OH transition at 15627\AA, 
these were also analyzed in a recent study of 
K and M giants of the Galactic bulge 
(Cunha \& Smith 2006). 
These OH lines analyzed in bulge red giants gave generally 
consistent oxygen abundances when compared to the results from   
the optical study by Fulbright et al. (2006) which analyzed the forbidden 
neutral oxygen ([OI]) in the same bulge stars.
In the more luminous Galactic Center stars studied here, however, 
the stronger OH lines become, in some 
cases, extremely strong and therefore very sensitive 
to the microturbulence parameter. 
Such dependence on microturbulence renders these 
lines too uncertain to be useful
as abundance indicators. The weaker OH line at 
15627\AA, however, could be measured
in all Galactic Center stars and served as the sole 
abundance indicator in IRS19 (which is
the most luminous Galactic Center star of the trio) and BSD114 (only 
observed at one H-band setting).

A simultaneous solution for CO was obtained, at the same 
time as OH was analyzed, from
computations of synthetic spectra of the first 
overtone vibration-rotation molecular 
lines of CO around $\lambda$23100\AA.  
The CO lines used are listed in Table 3.
The log gf-values and excitation potentials for 
the transitions were obtained from 
Goorvitch (1994). 
We note, again, that because the Galactic Center stars 
are more luminous when compared to the bulge 
giants, the stronger CO lines in the spectra
are so strong that they are extremely sensitive to 
the microturbulence parameter. 
The application of the simple requirement of searching 
for an agreement between the
abundances of strong and weak CO lines in 
the spectra would demand a higher microturbulent 
velocity, around 3 -- 4.5 km s$^{-1}$, and 
in some cases even higher. The formation of such strong CO
lines was investigated and it was found that the cores 
of these lines form extremely high
(i.e., at low values of optical depth) in the 
stellar atmospheres of these cool and 
luminous stars.  Such lines have large 
abundance uncertainties attached to them due to 
sensitivity to structure in the upper layers of 
the model atmosphere, for instance
the choice of the optical-depth boundary of the 
model atmosphere.  In this study the strong 
CO lines were rejected from the analysis and we relied on 
the weaker CO lines (from
Table 3) as abundance indicators.

Finally, the weaker CN lines in the H-band 
spectral region around 15540\AA\ (Table 3) were 
synthesized in order to derive nitrogen abundances 
for the Galactic Center stars.  This is the same 
spectral region (covered by some of the Phoenix spectra) that was 
analyzed in the bulge 
study by Cunha \& Smith (2006); that
paper is referred to for details on the  
construction of the CN line list. 
The CN lines are considerably weaker and somewhat 
more uncertain to measure 
in the cases of spectra with lower S/N.
They are not affected, on the other hand, by the same problems as 
found for the stronger OH and CO lines.

\subsubsection{Calcium and Iron}

Elemental abundances were 
derived from atomic lines of
Ca I and Fe I in addition to CNO.  
One suitable line of Ca I 
(at 22827\AA) and several Fe I lines were
used as abundance indicators, all in the K-band spectra
of the program stars. We also investigated the possibility of analyzing 
the Ti I transition 
at 15543\AA\ (which was previously studied in Cunha \& Smith 2006),
but this neutral titanium line was
extremely strong in all the observed spectra 
of the Galactic Center stars and 
therefore not useful in an abundance analysis. 

The sample of Fe I lines for which synthetic spectra were computed in this 
study is assembled in Table 3; 
this is the same sample as analyzed previously by Ramirez et al. (2000). 
The line list for the computation of synthetic spectra was 
the same as in Ramirez et al. (2000) 
with the only difference being that we derived 
new solar gf-values for all Fe I 
and Ca I lines. The solar
abundances adopted in the calculations were A(Fe)$_{\odot}$=7.45 and  
A(Ca)$_{\odot}$=6.30 (Asplund et al. 2005). The
derived solar gf-values (Table 3) were then checked via an analysis of 
the K-giant Arcturus, yielding 
results consistent with published abundances for this 
star (e.g. Smith et al. 2000). 
Use of the $\lambda$22827\AA\ Ca I
line with the solar gf-value yielded an abundance 
of A(Ca)=5.86 for $\alpha$ Boo, 
compared to the average
of 5.75 $\pm$ 0.15 from Smith et al. (2000), who used a set of optical Ca I
lines with laboratory gf-values.  
  
\subsection{Abundance Uncertainties}

Table 4 presents the abundance results for the studied elements 
C,N,O, Ca and Fe, while in Table 5 are summarized 
the corresponding uncertainties in the obtained abundances. 
The abundance uncertainties were calculated by modifying the 
stellar parameters of the model atmospheres by amounts corresponding to
what are expected to be maximum uncertainties in the 
effective temperatures ($\delta$ T$_{\rm eff}$= $\pm$ 200K); 
surface gravities ($\delta$ log g = $\pm$0.3 dex) 
and microturbulent velocities 
($\delta$ $\xi$ = $\pm$0.3 km s$^{-1}$).
Each one of these parameters was varied independently, 
adopting as a starting
model atmosphere one with T$_{eff}$= 3600K; log g=0.5 and 
microturbulence ($\xi$)= 2.2 km s$^{-1}$.
The total uncertainty for each element (presented in column 5 of Table 5) is 
the sum in quadrature of the individual errors. 
The sensitivities are different for the different lines, resulting from
differing typical line-strengths, different ionization or dissociation
energies, and different excitation energies.
The combination of all errors in Table 5 gives maximum uncertainties
of about 0.15 dex for calcium
and iron, around 0.2-0.25 for carbon and nitrogen, and around 0.3 dex 
for oxygen.

In addition to the sources of uncertainties discussed above, 
it is important to note that
non-LTE effects in the Galactic Center stars have not been evaluated 
and, therefore, the abundances derived here may suffer from systematic 
errors due to non-LTE effects. For the molecules studied here, especially
for the CO and OH vibration-rotation transitions, the non-LTE departures 
are expected to be 
small due to a combination of small transition probabilities (i.e., longer
lifetimes), which allow for thermalizing collisions, and the dilute
UV radiation fields in cool supergiants.
In the case of the atomic lines from 
Fe I and Ca I, their excitation potentials
are high enough and the depth of formation is deep enough such that non-LTE
effects should not be large.  

There is evidence in the abundance results suggesting that
non-LTE effects are probably not large.
The four disk M-giants analyzed for Fe and Ca yield values of [Ca/Fe] and
[Fe/H] that fall closely along the disk relation derived from other types
of stars (see the discussion in 4.3 and Figure 7); this would not be
expected if there were measurable effects from non-LTE.  In the case of
oxygen, the disk S-star HD172804 was observed in the OH lines and an 
oxygen abundance of A($^{16}$O)=8.55 was derived, in good agreement
with its previously published value of 8.50 (Smith \& Lambert 1990).  
Finally,
both oxygen and calcium are produced mostly in SN II and the abundance
ratio of Ca/O will be similar for similar metallicity SN II.
Since Ramirez et al. (2000) find that
the Galactic Center iron abundances are close to
solar, the values of Ca/O in near solar-metallicity stars
might be expected to be mirrored in the Galactic Center stars.
Indeed this
turns out to be the case; the average value of [Ca/O] in the five Galactic
Center stars for which both calcium and oxygen were measured is
[Ca/O]=+0.06$\pm$0.13.  This can be compared to the recent results from
Reddy et al. (2003) for 56 near-solar metallicity F and G dwarfs of 
[Ca/O]=-0.03$\pm$0.11.  Such close correspondence in very different types
of stars is not a coincidence and demonstrates that the abundances derived
here for Galactic Center stars are not subject to large non-LTE
effects (or other unspecified physical processes). 

\section{Discussion}

With the assembled set of elemental abundances from Fe and Ca in all program
stars, along with the $^{12}$C, $^{14}$N, and $^{16}$O 
abundances in a subset of stars,
the abundance distributions can begin to provide initial insights into both 
the nature of the evolved stars themselves (internal stellar evolution) and
the type of chemical evolution that has taken place in the inner central
60 pc of our Galaxy.
This discussion will focus first
on the Fe abundances, followed by C, N, and O taken together, and then both
O and Ca as typifying the $\alpha$-elements.  With an initial picture of the
nature of the chemical evolution in the Galactic Center in-hand, the abundance
distributions are examined in light of Galactic disk abundance gradients as
extrapolated into the center. 

\subsection{The Iron Abundance Distribution in the Galactic Center}

Iron is produced in both SN Ia and SN II and 
historically has been associated with
the overall metallicity of a stellar population.  
Its abundance is also well-defined
here by half-a-dozen lines, so this discussion begins with Fe.
The Fe results obtained in this study are fully
consistent with those derived by Ramirez et al. (2000) and Carr et al. (2000).
The average Fe abundance obtained for the 
sample of 10 stars is slightly higher (by $\sim$0.10--0.15 dex) than
the solar abundance of A(Fe)=7.45 recommended by Asplund et al (2005). 
The mean value is A(Fe)=7.59$\pm$0.06, where the standard deviation 
represents the scatter among the line-by-line values and 
probably underestimates the true uncertainties in the derived Fe abundances 
(the abundance uncertainty estimated for iron is in Table 5 and is roughly
0.15 dex).  
This value is quite a bit larger than the 
actual scatter in the individual abundances,
showing that systematic errors dominate the error budget.

Although a relatively small number of stars are 
analyzed in this study, we note the small abundance range found for iron
(from 7.51 to 7.67 in A(Fe)). This small scatter is in line 
with the narrow [Fe/H] distribution measured 
previously by Ramirez et al. (2000) for the same sample. 
It is much smaller, however, than the spread of more than 1.0 dex found for
the metallicity distribution of the older bulge giant population
(see e.g. the histogram shown in Figure 8 of 
Fulbright et al. 2006; Zoccali et al. 2003). 
The hypothesis that the Galactic center stars are drawn from the bulge 
metallicity distribution is rejected with P = $\sim$0.001 
using a K-S test. 
An additional difference between the bulge population and the Galactic 
center stars sampled here is the luminosity. 
Frogel \& Whitford (1987) find very few bulge stars 
with M$_{bol}$ brighter than $-4.2$; the stars
studied here range from
M$_{bol}$ = -4.5 to M$_{bol}$ = -9.0.  
The Galactic Center stars studied here were selected to provide 
a snapshot of current conditions in the Central Molecular Zone; 
they sample $R_G$ = 0.2 -- 30 pc
and have ages in the range 10 Myr to 1 Gyr.
It is not yet clear if the narrow Fe-abundance 
distribution is indicative of a
narrow distribution over a larger volume of space or time in the Galactic
Center, but this can be tested with future abundance studies 
of the Galactic Center
and bulge populations.

\subsection{Carbon, Nitrogen and Oxygen and the 
Evolutionary State of the Galactic Center Stars}

The CNO elemental trio are all very abundant and are important for
studying both chemical and stellar evolution.  
Three of the program stars were observed across sufficient wavelength 
intervals such that the most important CNO nuclei could be analyzed. 
The studied nuclei include $^{12}$C, $^{14}$N and $^{16}$O which
are the dominant CNO species. Carbon-13 is not included nor is $^{15}$N, 
although in terms of numbers
these are relatively minor species. As $^{12}$C is depleted to levels 
below $^{14}$N (with $^{14}$N being built-up during the CN-cycle),
the expected $^{12}$C/$^{13}$C CN-cycle equilibrium ratio of 3.5 will be 
reached such that N($^{13}$C) $<$ N($^{12}$C) $<$ N($^{14}$N),
so lack of a carbon-13 abundance will not hinder our conclusions about
chemical evolution significantly. Nitrogen-15 is expected
to be a very minor constituent so its contribution to the total abundances 
is not important for our discussion.

The program stars observed here are almost certainly mixed to some degree, 
with some fraction of their convective
envelope material having been exposed to CN-cycle H-burning, or possibly
ON-cycle H-burning. This scenario can be 
tested using the respective CNO abundances.
If the convective envelopes have been subjected to 
CN-cycle proton captures, then
the total sum of $^{12}$C, $^{13}$C, and $^{14}$N 
will be approximately conserved 
and this can be probed by plotting
A($^{14}$N) versus A($^{12}$C), as shown in the top panel of Figure 6. 
Carbon-13 is ignored as we have
no information on it, but $^{13}$C will 
not significantly alter any conclusions obtained
from Figure 6.  
The solid circles with errorbars are the $^{12}$C and $^{14}$N abundances 
for Galactic Center stars IRS19, BSD72 and BSD124 and the 
results for IRS7 by Carr et al. (2000)
are represented by the red square.
The solar point is plotted with the straight line delineating a solar 
$^{12}$C/$^{14}$N ratio scaled 
in metallicity. The solid curves are defined by constant abundances of 
$^{12}$C + $^{14}$N; the two
curves show carbon plus nitrogen abundances scaled-up from solar by 0.1 
and 0.2 dex, respectively.

All three stars studied here are well-fit by slightly elevated
metallicities (as found from the Fe abundances) 
and CN-cycled material in their atmospheres
(somewhat depleted $^{12}$C and elevated $^{14}$N). 
Also shown in the two panels of Figure 6
are the $^{12}$C and $^{14}$N abundances for 
IRS7 from Carr, Sellgren \& Balachandran (2000). 
IRS7 is considerably more massive and luminous than IRS19, BSD72, or 
BSD124 and shows a much lower $^{12}$C
and much larger $^{14}$N abundance than any of the three 
lower luminosity stars. The $^{14}$N
abundance in IRS7 is so large that it may indicate that 
the CN-cycle alone cannot account for such a large amount of 
nitrogen. Indeed, Carr et al. (2000) find a very low
$^{16}$O abundance and argue that this is evidence 
of the presence of ON-cycle material
in the atmosphere of IRS7 (where $^{16}$O has been 
converted to $^{14}$N). Being much more massive and luminous 
(it is the brightest star in the Galactic Center 
at H and K) IRS7 is presumably more deeply mixed 
and this suggestion is borne out by the lower panel of Figure 6 where
we show log (N($^{16}$O)/N($^{14}$N)) versus M$_{bol}$ for 
the four Galactic Center stars.

\subsection{$\alpha$-Element Abundances in the Galactic Center}

Determination of [$\alpha$/Fe] in any stellar population and, 
in particular, in 
young stars in the Galactic Center provides important information 
on chemical
enrichment, gas infall, and gas outflow. 
One possible scenario is that the relatively young ages 
(1 - 100 Myr) of the brightest stars in the 
Galactic Center support the idea that the Galaxy's central bar has 
driven disk gas into the Galactic 
Center to fuel star formation throughout the Galaxy's 
history, continuously or in bursts of star 
formation (Serabyn \& Morris 1996; Morris \& Serabyn 1996; 
Blum et al. 2003; Figer et al. 2004). In particular,
the star formation conditions in the Galactic Center are 
very different from 
those in the solar neighborhood, and are predicted to 
result in an initial mass function 
(IMF) weighted toward massive stars (Morris 1993).
Furthermore, models of supernova enrichment in an 
environment dominated by massive stars predict a high relative abundance of 
$\alpha$-elements compared to Fe, [$\alpha$/Fe] (e.g., Wheeler et al. 1989). 

The results presented in Figure 6 show the presence of CN-mixing
in the three evolved stars but not the very deep mixing that is observed in
the much more luminous red supergiant IRS7. 
This suggests that oxygen has been
effectively untouched by deeper mixing in these three
stars.  Also, because the net
result of ON-cycle mixing would be to reduce the oxygen 
abundance, the oxygen results
in fact represent a lower limit on the initial oxygen abundance
for the original gas from which these stars formed.
The average oxygen abundance found for our sample of five stars 
is $<$A(O)$>$=9.04 $\pm$ 0.19. The scatter observed in the sample oxygen abundance 
distribution can be explained by the uncertainties in the analysis 
(discussed in Section 3.4).

The values of [O/Fe] in the Galactic Center stars can be compared 
to other samples from the Galaxy.
A comparison to the Galactic disk and halo is shown in the top panel
of Figure 7. In the five stars studied here for oxygen it is apparent that
the average value of [O/Fe] is elevated relative to the disk: $<$[O/Fe]$>$ = +0.22 $\pm$ 0.15. 
Both the narrow [Fe/H] and elevated $\alpha$-element abundances are 
reminiscent of the results obtained by Rich \& Origlia (2005) for
M-giants in the bulge 500 parces from the galactic center.
This behavior of enhanced values of [O/Fe] at [Fe/H] $\ga$ 0.0
is also similar to what has been found by a number of different 
studies of old K and M red-giants in the Galactic bulge 
(Cunha \& Smith 2006; Zoccali et al. 2006; Fulbright et al. 2006).
The interpretation of the bulge pattern is that the
chemical enrichment was rapid and dominated by SN type II to quite high
Fe abundances. 
The population of stars studied here, however, is distinct 
and very different from the bulge, as the Galactic Center 
stars are all rather young and massive (their estimated
masses are presented in Table 1). 
Additionally, the K luminosity function of the bulge only 
contains stars with $K_0$ $\ga$ 8.0 (DePoy et al 1993; Zoccali et al. 2003) 
while the Galactic Center stars observed here have $K_0$ = 3.0 -- 7.4.
 
This study also includes an additional $\alpha$-element, calcium.  The
results for calcium, in the form of [Ca/Fe] versus [Fe/H], are shown in the
bottom panel of Figure 7.  The small open symbols are results for
Galactic field disk and halo from a number of studies noted in the 
figure caption. The large filled symbols are
the [Ca/Fe] results for the Galactic Center stars.  Because of the
larger number of Galactic Center stars with measured Ca abundances, the
elevated values of [Ca/Fe] at large metallicity are even more striking
than the results for [O/Fe].  Of further interest are the results for
[Ca/Fe] in the four disk M-giants analyzed in the same way as the Galactic
Center; these are plotted as the filled squares.  The disk M-giants
exhibit quite small scatter around the disk trend of [Ca/Fe] versus [Fe/H].
Calcium behaves in a similar way as oxygen, relative to Fe. The average
[Ca/Fe] abundance for our Galactic Center sample is $<$[Ca/Fe]$>$= +0.34 $\pm$ 0.14:
here again the abundance spread can be explained by the errors in the analysis. 
A similar behavior for the two elements oxygen and calcium provides strong 
evidence that this sample of Galactic Center stars
are composed of chemical compositions containing larger abundances of the
$\alpha$-elements relative to Fe when compared to the Galactic disk at the
same metallicity. 

The results derived here of elevated O and Ca abundances, relative to Fe, 
are unexpected. It is therefore of interest to carefully evaluate if
it is possible that reasonable changes in the adopted stellar parameters
could erase the observed $\alpha$-enhancements.
The various abundance sensitivities to the different
stellar parameters are listed in Table 5. Examination of this table
reveals that the only plausible parameter that could be changed in
order to drive the elevated values of [$\alpha$/Fe] to near solar
would be the effective temperature; a change of $\delta T_{eff}$ $\sim$ - 400K 
would be needed in order to force an average [Ca/Fe] $\sim$ 0.0 for
this sample of stars. Such a systematic shift in the T$_{eff}$-scale
is unlikely. Inspection of Figure 3 shows that 
such a change would shift $T _ {\mbox eff}$ for
these Galactic Center stars to values of $\sim$ 2800 -- 3500 K. These lower
temperatures would produce strong H$_2$O absorption for the coolest stars, which
is not observed (Blum et al. 2003), and would put the the highest luminosity 
stars in extreme disagreement with the stellar model tracks (Fig. 4). 
All this combined seems to indicate that the stellar members of the Central 
and Quintuplet Clusters analyzed here contain larger than solar ratios of 
Ca/Fe and O/Fe. 

\subsubsection{Comparisons with Other Abundance Measurements 
in the Galactic Center}

Independent measurements of the abundances of cool, massive stars 
in the Galactic Center are offered by studies of hot, massive stars 
in the three massive clusters in the central 100 pc, and by
studies of the gas in the Central Molecular Zone.
The massive young stars (formed millions of years ago) and the 
present-day gas in the central 200 pc both provide snapshots of how 
far the Galactic Center has chemically evolved to date.
One important aspect of such comparisons is the fact that very 
different analysis techniques are used to measure nebular
abundances and chemical compositions of lines in the atmospheres of stars.

Najarro et al. (2004) measure the N abundance from near-infrared spectra of
WNL stars in the Arches Cluster (projected $R_G$ = 24 pc) in the Galactic 
Center.  They find best agreement between observations
and theory for solar metallicity stellar evolution models.
Najarro (2006) presents preliminary
models of the near-infrared spectra of two stars in 
the Quintuplet Cluster, and finds a solar Fe abundance for both.
Najarro (2006) also finds initial results that the $\alpha-$elements
Si and Mg are enhanced by $\sim$0.1--0.3 dex over solar.

Infrared spectroscopy has probed two H II regions in the Galactic Center, 
G359.98-0.08 and G0.095+0.012 (ionized by the Arches Cluster),
at projected $R_G$ = 6.6 and 24 pc.
This reveals enhanced $\alpha-$elements,
[O/H] = 0.1--0.6 and [S/H] = 0.2--0.4, and also [N/H] = +0.5--0.8
(Simpson et al. 1995;
Afflerbach et al. 1997;
Rudolph et al. 2006).
Shields \& Ferland (1994) model the far-infrared spectroscopy of Sgr A West,
the \ion{H}{2} region surrounding the Central Cluster.
They find a best fit for [Ar/H] = +0.3 and [Ne/H] = 0.0, and
conclude that the overall metallicity is between solar and twice solar.
Lutz et al. (1993) measure [Fe/H] $\geq$ -0.1 in the ``Mini-Cavity'' 
region of Sgr A West, and point out that unless there is absolutely 
no depletion of Fe onto grains, highly supersolar Fe abundances are required.
They suggest that fast shocks from stellar winds have destroyed dust grains
and released Fe into the gas.

Recent X-ray measurements of the gas-phase Fe abundance in the 
Galactic Center have been made.
Sgr A East has a mean value of [Fe/H] = 0.1 (Sakano et al. 2004) to
0.6 (Maeda et al. 2002).
Sgr A East is a young ($\sim$10$^4$ yr old) supernova remnant which 
is physically interacting with the Sgr A West \ion{H}{2} region
(Yusef-Zadeh \& Morris 1987).
Evidence suggests that the Sgr A East supernova shock front passed through
Sgr A West $\sim$300 yr ago (Maeda et al. 2002).
Sakano et al. (2004) measure [Ca/H] = +0.4, [S/H] = +0.3, [Si/H] = +0.7,
and [Ar/H] = +0.1 in Sgr A East. 
Sgr A East, however, is expected to be self-enriched, and it is 
under debate what type supernova it was.
Koyama et al. (2007) find [Fe/H] = +0.5 from the absorption column 
density of Fe nuclei towards the Galactic center; they argue that
most of this absorption is from dense clouds local to the Galactic Center.
The absorption towards the Galactic Center, however, is well-known to have
significant contributions from foreground material (Magnani et al. 2006).

These results on infrared spectroscopy of hot massive stars, infrared 
spectroscopy of H II regions, and X-ray spectroscopy of Sgr A East
and the diffuse gas in the Galactic center, offer an independent
assessment of the chemical abundances in the Galactic Center.
We conclude from the discussion above that there is independent evidence for an enhancement of 
$\alpha$-elements of 0.2--0.4 dex in the Galactic Center, from 
O and S in the gas, and from Si and Mg in hot massive stars.
The X-ray Fe abundances are difficult to interpret, but Najarro (2006) 
find [Fe/H] = 0.0 in hot massive stars, as we do in cool massive stars.

\subsection{Connections with Metallicity Gradients of the Galactic Disk}
 
It is of interest to compare the abundances of stars
located within 60 pc of the Galactic Center with the abundances
that have been measured further out in the Galactic disk in
order to examine the possibility that gas
has been driven to the Galactic Center, for example,
by the presence of the central bar.
This comparison can be done via a discussion of 
the metallicity gradients derived 
for the more distant parts of the Milky Way 
and centers around the question of 
whether these gradients smoothly connect disk abundances to 
those in the center of the Galaxy. 
As discussed in Smartt et al. (2001) 
there seem to be indications of differences between the
inner disk and the Galactic Center which are 
inferred from the general lack of HII 
regions, molecular gas and young stars in the region 
between the central clusters and 
Galactic radii corresponding to $\sim$3.5 kpc. 

Stellar abundances as a function of galactocentric distance 
are illustrated in Figure
8 for the elements oxygen, calcium and iron. 
The data are taken from a number of
studies as indicated in the figure. 
The top panel shows [O/H] versus R$_{g}$
with disk abundances coming from both OB stars and Cepheids.
The Cepheids are much cooler than the OB stars with oxygen abundances in the
Cepheids derived from neutral
oxygen lines, while in the OB stars O II lines are analyzed. 
Despite these different abundance indicators, 
values of O/H for the two groups of stars largely overlap
and show a gradual increase of oxygen with 
a decrease in R$_{g}$ (e.g. Daflon \& Cunha fit a linear
slope of -0.04 dex/kpc for their OB star data). 
Given the scatter in the data points when 
considering both type of stars, the elevated 
values of [O/H] found here in the Galactic 
Center stars do not seem to be at odds with a gradual and modest increase 
in the oxygen abundances for R${_g}$ close to zero.
 
Although calcium and iron cannot be studied in OB stars, 
both elements are contained in 
the more extended list of elements whose abundances 
can be derived from spectroscopy of 
Cepheids. The middle panel of Figure 8 shows 
[Ca/H] as a function of galactocentric
distance. 
As with oxygen there is a gradual increase in the Ca abundance 
with smaller values
of R${_g}$. 
In this case, however, there are no Cepheid measurements inside of roughly
5 kpc, so there is a large gap in any connection to the Galactic Center. 
Because of the lack of Ca abundances in the direction of the 
Galactic Center it would be more difficult to
argue that the values of [Ca/H] found in the stars 
there represent a smooth extension
of the trend observed in the disk.

Concerning iron, Luck et al (2006), 
who summarize all of their group's previous
studies, find that the overall gradient for the Cepheid
sample is -0.068 dex/kpc for galactocentric distances 
ranging between roughly 4 and 16 kpc.
A simple extrapolation of this iron gradient to  
R$_{g}$=0 (Galactic center), 
would yield an iron abundance that is enhanced 
by +0.6 dex relative to the solar value 
(or, A(Fe)=8.1): this abundance is significantly 
higher than the average 
Fe abundance obtained in this study for the 
Galactic center stars ($<$A(Fe)$>$=7.6).
This is well-illustrated in the bottom panel of Figure 8. 
In this case, the gradient
in the Fe abundance is measurably steeper than that observed for calcium. 
In order for
the Galactic Center stars to have Fe abundances that 
are representative of the inner
disk would require that this trend flattens markedly 
at small values of R${_g}$.

The hint that there may be different gradients in the Ca and Fe abundances can
be probed further by examining the ratio of calcium to iron as a function of
R$_{g}$: these results are shown in Figure 9.  It is clear that there is a
general decrease in [Ca/Fe] as R$_{g}$ decreases, whereas the Galactic Center
stars contain elevated values of [Ca/Fe].  In order for the elemental
abundance distributions of the inner disk (R$_{g}$ $\sim$ 3-4 kpc) to connect to
the Galactic Center would require a significant upturn in the trend of
[Ca/Fe].  This could occur as a result of either a rapid increase in [Ca/H] or
a rapid decrease in [Fe/H], or some combination of both.  A re-examination 
of Figure 8 might allow for a
turnover in [Fe/H] below R$_{g}$ $\sim$ 4 kpc, 
however there are no iron abundance
data points in this part of the disk.  It is perhaps safer to conclude that  
the significantly enhanced values of [Ca/Fe] found in all
10 Galactic Center stars studied here do not seem to be consistent with simple
extrapolations of disk abundance gradients into the center. 

The high value of [Ca/Fe] observed in the
Galactic Center stars might argue against gas
from the disk, which has roughly a solar value of [Ca/Fe],
reaching the Galactic Center.
Similiarly, the narrow [Fe/H] distribution observed in the
Galactic Center stars might argue against gas
from the bulge, which has a broad [Fe/H] distribution, being
a major contributor to the present-day abundances in the
Galactic Center.
Disk gas driven into the Galactic Center by the bar,
however, and infall of gas from stellar mass loss in the bulge,
are both believed to fuel sustained star formation
in the Galactic Center
(Serabyn \& Morris 1996; Morris \& Serabyn 1996;
Stark et al. 2004).
Furthermore, we observe enrichment of the Galactic
Center gas in the present day, in the
locally high abundances observed for the supernova remnant
Sgr A East (Sakano et al. 2004).
The unique abundance pattern in Galactic Center stars
may reflect the variety of contributors to the
chemical evolution of this regions.

\subsection{Implications for Chemical Evolution in the Galactic Center}

Chemical evolution models for the specific case of the
central 100 pc of the Galactic Center, which take into account its 
extended star formation history as well as its various sources of gas infall
and possible mechanisms for gas outflow are not available in the
literature. Such detailed modeling would be extremely welcome for the 
interpretation of existing and future Galactic Center abundance results. 
In lieu of such detailed models, our abundance results can be considered 
in a more simplified discussion of chemical evolution.

The evidence presented here of large values of [O/Fe] and [Ca/Fe] in Galactic
Center stars points to the dominant role played by SN II chemical enrichment
in the gas from which these particular stars formed. 
This result is not expected from simple star formation history (SFH) and 
chemical enrichment models. For example, Figer et al. (2004) found that 
the Galactic Center luminosity function could be fit by a continuous star 
formation history (see Blum et al. 2003 for a more general treatment which 
does not lead to a continuous SFH). 
In a continuous SFH (or a SFH with repeated bursts of star formation), SN Ia will begin to contribute,
after $\sim$1 Gyr, large amounts 
of Fe and drive the [O/Fe] and [Ca/Fe] values down towards ones more 
solar-like in the younger stars; such solar-like ratios are not observed here.

There are solutions that would reconcile the picture of continous star
formation with the enhanced values of [$\alpha$/Fe] observed here in
young Galactic Center stars. One could invoke either an IMF that is
weighted towards more massive stars, or appeal to the stars' youth
and their restricted location in space. In a top-heavy IMF, the larger
fraction of massive stars would overproduce both oxygen and calcium
when compared to disk values. In this case, the Fe produced by the 
lower mass SN Ia systems would not be enough to yield values of
[$\alpha$/Fe]=0 in the gas from which the young stars formed. The second
scenario mentioned above would argue that the star formation process
that has produced the masssive clusters near the Galactic Center could
have self-polluted their local region on short time-scales via SN II
within the clusters themselves. Continuing star formation within the
cluster environment would then occur in gas enriched in calcium and
oxygen. This scenario would mean that the elevated values of 
[O/Fe] and [Ca/Fe] are peculiar to the time and place occupied by
the massive clusters. Olive and Schramm (1982) suggested that such
an effect may occur in large regions of active star formation. 
An analysis of a more extended sample of Galactic Center stars could
potentially test the two scenarios discussed above.

\section{Conclusions}
 
The 10 cool, luminous, and evolved Galactic Center stars analyzed here
present nearly uniform iron abundances that are slightly larger
than solar, with a mean and standard deviation of A(Fe)=7.59$\pm$0.06
(or [Fe/H]=+0.14).  This value for Fe confirms the near-solar, or
slightly elevated values reported previously in Galactic Center stars
by Ramirez et al. (2000) and Carr et al. (2000).  The derived small 
scatter in the Fe abundances is within the uncertainty of an individual
star's abundance, thus a single-valued iron abundance typifies the
particular Galactic Center population sampled by these 10 stars.
Similarly, the oxygen and calcium abundances for the studied sample
can also be represented by single abundance values within the expected
uncertainties in the analysis.

Analysis of the specific CNO isotopes $^{12}$C, $^{14}$N, and $^{16}$O
reveals the presence of CN-cycle material having been mixed into the
atmospheres of these evolved stars; that is, $^{14}$N is significantly
enhanced, accompanied by a slightly lower $^{12}$C abundance, such
that the sum of $^{12}$C and $^{14}$N nuclei is roughly constant, as
expected for CN-cycle mixing.  The 3 stars studied for CNO fall along
so-called CN-mixing lines in a $^{14}$N versus $^{12}$C diagram for an
initial $^{12}$C/$^{14}$N value that is about solar.  

Perhaps the most significant result of this study is the $\alpha$-enhanced 
elemental abundances found for the Galactic Center stars. The [O/Fe] and
[Ca/Fe] values are on average  +0.2 -- +0.3 dex higher than typical disk values
at similar Fe abundances. 
These abundance patterns are consistent with a larger fraction of SN II
relative to SN Ia when compared to the disk. A reason for this might be an IMF for
the Galactic Center weighted towards more massive stars, or recent, local
SN II chemical enrichment within the central 50 pc of the Galaxy.

\acknowledgements

K.C. thanks Ken Freeman for discussions.
K.S. thanks Marc Pinsonneault and Rick Pogge for discussions.
Based on observations obtained at the Gemini Observatory,
which is operated by the Association
of Universities for Research in Astronomy, Inc., under a cooperative
agreement with the NSF on behalf of the Gemini partnership: the National
Science Foundation (United States), the Particle Physics and Astronomy
Research Council (United Kingdom), the National Research Council (Canada),
CONICYT (Chile), the Australian Research Council (Australia),
CNPq (Brazil), and CONICRT (Argentina).  This
paper uses data obtained with the Phoenix infrared spectrograph,
developed and operated by the National Optical Astronomy Observatory.
Based on observations at the Infrared Telescope Facility, which is operated 
by the University of Hawaii under Cooperative Agreement no. NCC 5-538 with 
the National Aeronautics and Space Administration, Science Mission 
Directorate, Planetary Astronomy Program.
This work is also supported in part by the National Science Foundation through
AST02-06331 (KS, SVR) and AST06-46790 (KC, VVS), and NASA through 
JPL Agreement 1265550 (KS) and NAG5-9213 (KC, VVS).

%\clearpage

\clearpage

\begin{deluxetable}{cccccccc}
\setcounter{table}{0}  
\tablewidth{0pt}  
\tablecaption{Galactic Center Stars and Stellar Parameters}
\tablehead{
\colhead{Star} &   
\colhead{T$_{\rm eff}$(K)} &
\colhead{Log g\tablenotemark{a}} &
\colhead{$\xi$\tablenotemark{b}} &
\colhead{$\zeta$\tablenotemark{b}} &
\colhead{M$_{bol}$} &
\colhead{Mass(M$_{\odot}$)} &
\colhead{R$_{g}$\tablenotemark{c}}
} 
\startdata 
BSD72  &   3880 &  0.8 &  2.1 &   10-12 & -4.54 & 5   & 1.6  \\
BSD114 &   3150 & -0.3 &  2.6 &   12-16 & -5.27 & 2   & 2.1  \\ 
BSD124 &   3735 &  0.4 &  2.5 &   12-14 & -5.50 & 6.5 & 1.9  \\ 
BSD129 &   3800 &  0.5 &  2.5 &   9-14  & -5.33 & 6.5 & 2.2  \\ 
BSD140 &   3200 &  0.0 &  3.1 &   14-16 & -4.62 & 2   & 2.0  \\ 
IRS7   &   3650 & -0.5 &  3.2 &   25    & -9.00 & 22  & 0.21 \\ 
IRS11  &   3625 &  0.3 &  2.3 &   11-15 & -5.04 & 4.5 & 0.62 \\ 
IRS19  &   3850 &  0.1 &  3.0 &   13-15 & -7.28 & 15  & 1.0  \\ 
IRS22  &   3750 &  0.2 &  2.3 &   11-16 & -6.49 & 10  & 1.0  \\ 
VR5-7  &   3600 & -0.15&  2.6 &   14-16 & -7.62 & 14  & 31.0  \\ 
\enddata
\tablenotetext{a}{Units of cm s$^{-2}$}
\tablenotetext{b}{Units of km s$^{-1}$} 
\tablenotetext{c}{Projected galactocentric distance in pc.}
\end{deluxetable}

\clearpage

\begin{deluxetable}{ccccccc}
\setcounter{table}{1}
\tablewidth{0pt}
\tablecaption{Disk Stars and Stellar Parameters}
\tablehead{
\colhead{Star} &
\colhead{T$_{\rm eff}$(K)} &
\colhead{Log g\tablenotemark{a}} &
\colhead{$\xi$\tablenotemark{b}} &
\colhead{$\zeta$\tablenotemark{b}} &
\colhead{M$_{bol}$} &
\colhead{Mass(M$_{\odot}$)}
}
\startdata
HD172804 & 3400 & 0.3  &  2.9 &   10    & -4.7 & 2.2 \\
HR6146   & 3250 & 0.14 &  2.1 &   10-12 & -4.1 & 1.5 \\
HR6702   & 3300 & 0.14 &  2.1 &   9-11  & -4.1 & 2   \\
HR7442   & 3450 & 0.09 &  2.2 &   8-12  & -4.4 & 2.5 \\
HR8062   & 3450 & 0.31 &  2.3 &   8-12  & -4.3 & 2   \\
\enddata
\tablenotetext{a}{Units of cm s$^{-2}$}
\tablenotetext{b}{Units of km s$^{-1}$}
\end{deluxetable}

\clearpage

\begin{deluxetable}{ ccc }
\setcounter{table}{2}
\tabletypesize{\footnotesize}
\tablewidth{0pt}
\tablecaption{Spectral Lines (analyzed via Spectrum Synthesis)}
\tablehead{ $\ \ \ \ \ \ \ \ \ \ $ $\lambda$(\AA) 
$\ \ \ \ \ \ \ \ \ \ \ \ $&
\multicolumn{1}{c} {$\ \ \ \ \ \ \ \ \ \ $ $\chi$(eV) 
$\ \ \ \ \ \ \ \ \ \ \ $} &
\multicolumn{1}{c}{ $\ \ \ \ \ \ \ \ \ \ $  log gf 
$ \ \ \ \ \ \ \ \ \ \ \ $ }}
\startdata
   Fe I   &       &          \\
22381.27  & 5.844 &  -1.511  \\
22386.90  & 5.033 &  -0.224  \\
22391.22  & 5.320 &  -1.625  \\
22398.98  & 5.099 &  -1.134  \\
22818.82  & 5.792 &  -0.992  \\
22838.60  & 5.099 &  -1.222  \\
  Ca I    &       &          \\
22827.283 & 4.624 &  -0.246  \\
$^{12}$C$^{14}$N & &         \\
15530.987 & 0.89 &   -1.519  \\
15544.501 & 1.15 &   -1.146  \\
15552.747 & 0.90 &   -1.680  \\
15553.659 & 1.08 &   -1.285  \\
15563.376 & 1.15 &   -1.141  \\
$^{16}$OH &      &           \\
15535.489 & 0.51 &  -5.233   \\
15560.271 & 0.30 &  -5.307   \\
15568.807 & 0.30 &  -5.270   \\
15572.111 & 0.30 &  -5.183   \\
15627.290 & 0.89 &  -5.435   \\ 
15627.292 & 0.89 &  -5.435   \\
15627.413 & 0.54 &  -5.198   \\ 
$^{12}$C$^{16}$O &      &    \\
23108.742 & 1.51 &  -4.907   \\ 
23121.469 & 1.55 &  -4.900   \\
23162.381 & 1.66 &  -4.879   \\
23191.951 & 1.74 &  -4.857   \\
\enddata
%\tablecomments{This page unintentionally left blank.}
\end{deluxetable}

\clearpage

\begin{deluxetable}{ cccccc }  
\setcounter{table}{3}  
\tablewidth{0pt}  
\tablecaption{Abundances\tablenotemark{a}}

\tablehead{  Star &   
\multicolumn{1}{c} {A($^{12}$C)} &
\multicolumn{1}{c} {A($^{14}$N)} &
\multicolumn{1}{c} {A($^{16}$O)} &
\multicolumn{1}{c} {A(Fe)} &
\multicolumn{1}{c} {A(Ca)} 
} 
\startdata 
BSD72    &   8.44 &  8.20  &  8.80  &  7.51 &  6.66 \\  
BSD114   &   --   &  --    &  8.95  &  7.57 &  6.60 \\
BSD124   &   8.45 &  8.48  &  9.30  &  7.61 &  7.00 \\
BSD129   &   --   &  --    &  --    &  7.67 &  6.76 \\
BSD140   &   --   &  --    &  --    &  7.52 &  6.50 \\
IRS7     &   --   &  --    &  --    &  7.66 &  6.86 \\
IRS11    &   8.40 &  --    &  9.00  &  7.53 &  6.60 \\
IRS19    &   8.54 &  8.33  &  9.15  &  7.63 &  6.66 \\
IRS22    &   --   &  --    &  --    &  7.57 &  6.93 \\
VR5-7    &   --   &  --    &  --    &  7.60 &  6.51 \\
         &        &        &        &       &       \\
HD172804 &   8.50 &  --    &  8.55  &  --   &  --   \\
HR6146   &   --   &  --    &  --    &  7.47 &  6.20 \\
HR6702   &   --   &  --    &  --    &  7.42 &  6.18 \\
HR7442   &   --   &  --    &  --    &  7.22 &  6.08 \\
HR8062   &   --   &  --    &  --    &  7.31 &  6.31 \\
\enddata
\tablenotetext{a}{A(X)= log[n(X)/n(H)] + 12.}
\end{deluxetable}

\clearpage
\begin{deluxetable}{ ccccc }
\setcounter{table}{4}
\tablewidth{0pt}
\tablecaption{Abundance Uncertainties}

\tablehead{ Element &
\multicolumn{1}{c} {$\delta$T=+200K} &
\multicolumn{1}{c} {$\delta$log g=+0.3} &
\multicolumn{1}{c} {$\delta \xi$=+0.3km s$^{-1}$} &
\multicolumn{1}{c} {$\Delta$\tablenotemark{a}}  
}
\startdata
$\Delta$ $^{12}$C & +0.02  & +0.14 & -0.12 &  $\pm$0.22 \\
$\Delta$ $^{14}$N & +0.20  & -0.11 & -0.05 &  $\pm$0.25 \\
$\Delta$ $^{16}$O & +0.30  & -0.05 & -0.12 &  $\pm$0.33 \\
$\Delta$ Ca       & +0.10  & +0.02 & -0.11 &  $\pm$0.15 \\
$\Delta$ Fe       & -0.04  & +0.07 & -0.11 &  $\pm$0.15 \\
\enddata
\tablenotetext{a}{[($\delta T)^{2}$ + ($\delta$log g)$^{2}$ + 
($\delta \xi)^{2}$]$^{1/2}$}
\end{deluxetable}

\clearpage

\begin{figure}
\vspace*{-8mm}
\epsscale{0.90}
\plotone{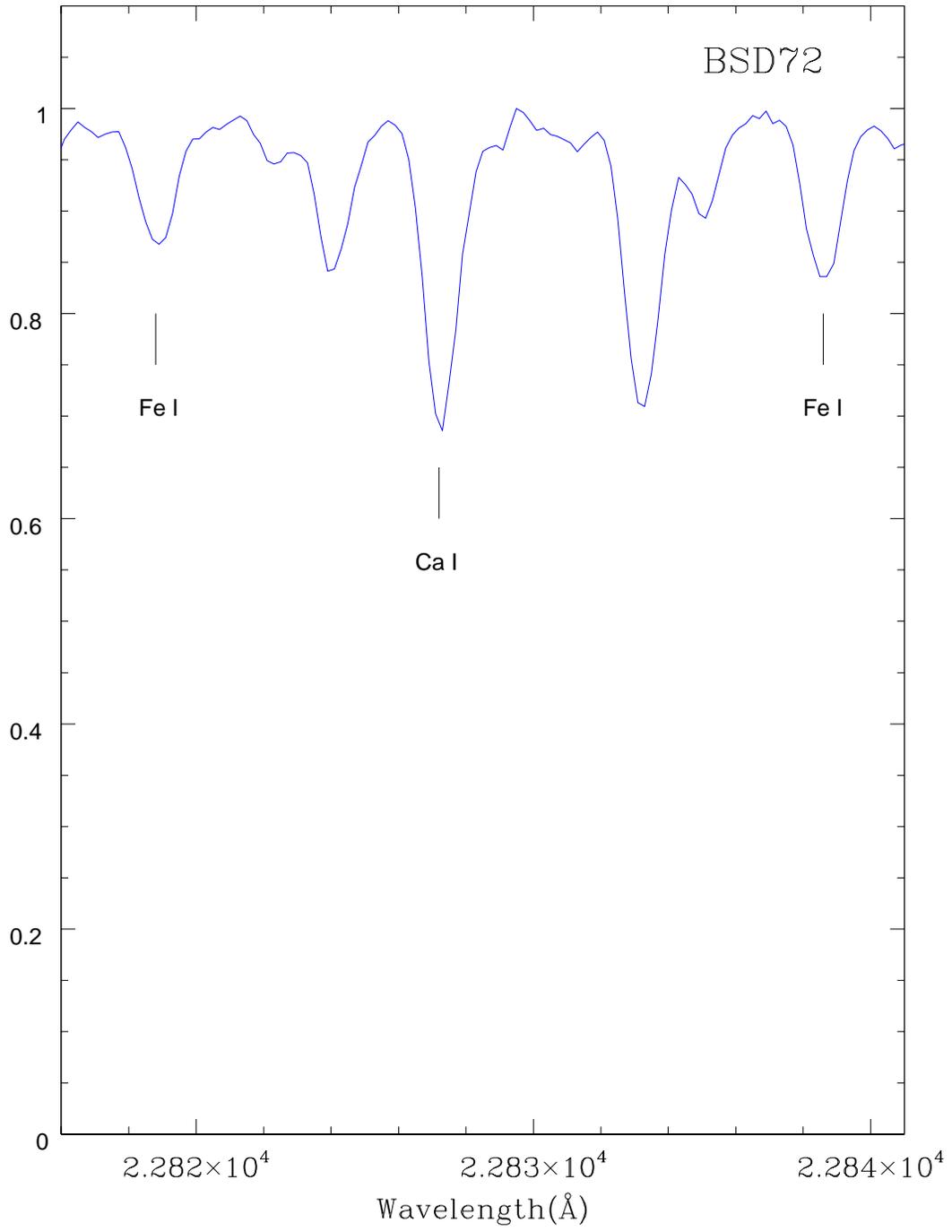}
\figcaption{Sample spectrum for Galactic Center
star BSD72 in the region containing two 
sample Fe I lines (at 22818.8 and 22838.6\AA) and the Ca I line 
(at 22827.3\AA) analyzed in this study.
\label{fig1}}\end{figure}

\begin{figure}
\epsscale{.72}\plotone{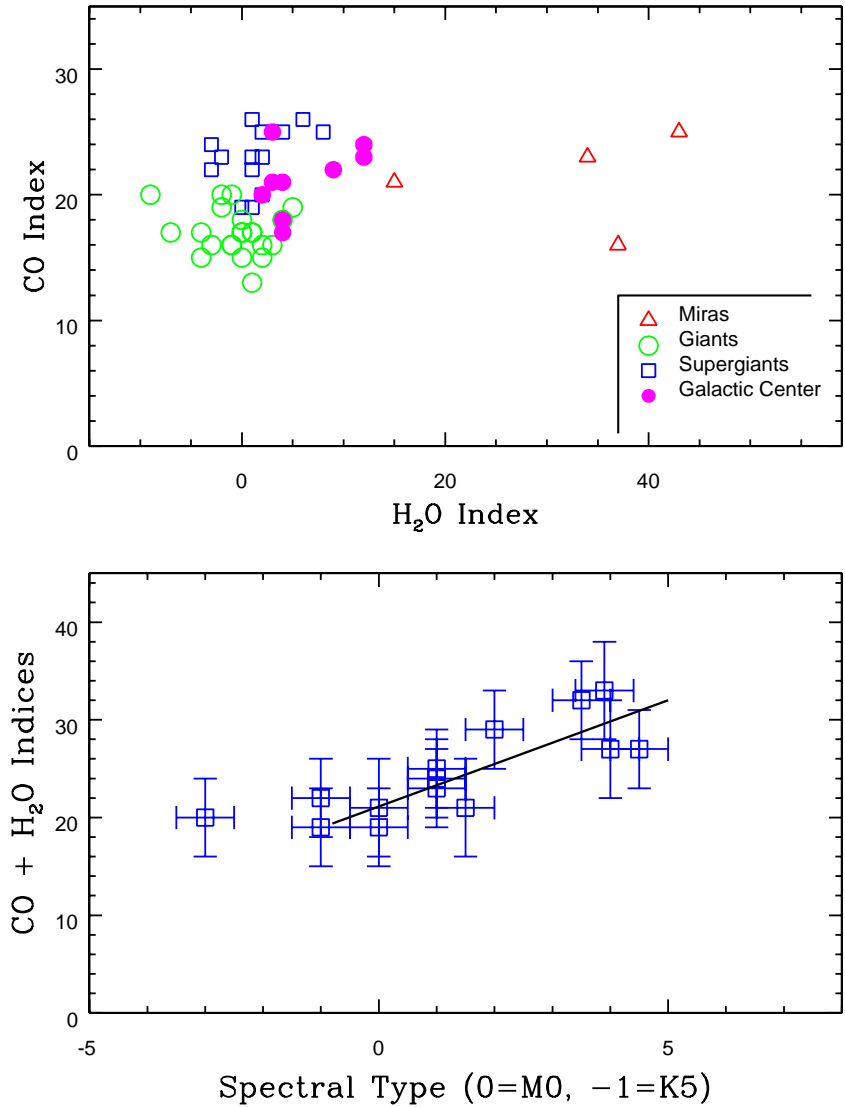} 
\figcaption{{\it Top panel}: CO and H$_{2}$O absorption indices in various types 
of cool luminous stars (from Blum et al. 2003).  The giants and supergiants segregate
quite cleanly in this diagram, with the CO index increasing with increasing
luminosity.  The Miras form an extended sequence towards extremely strong
H$_{2}$O absorption.  The Galactic Center stars fall closest to the more
luminous giants to supergiants, along a sequence defined by somewhat stronger
H$_{2}$O absorption.  {\it Bottom panel}: Total CO + H$_{2}$O absorption versus 
spectral type for supergiants (from Blum et al. 2003).
Between spectral types M0 to M5 there is a well-defined increase in CO + H$_{2}$O
towards later spectral types. The straight line is a linear fit to the data from M0 to M5.
\label{fig2}}\end{figure}

\begin{figure}
\epsscale{0.77}
\plotone{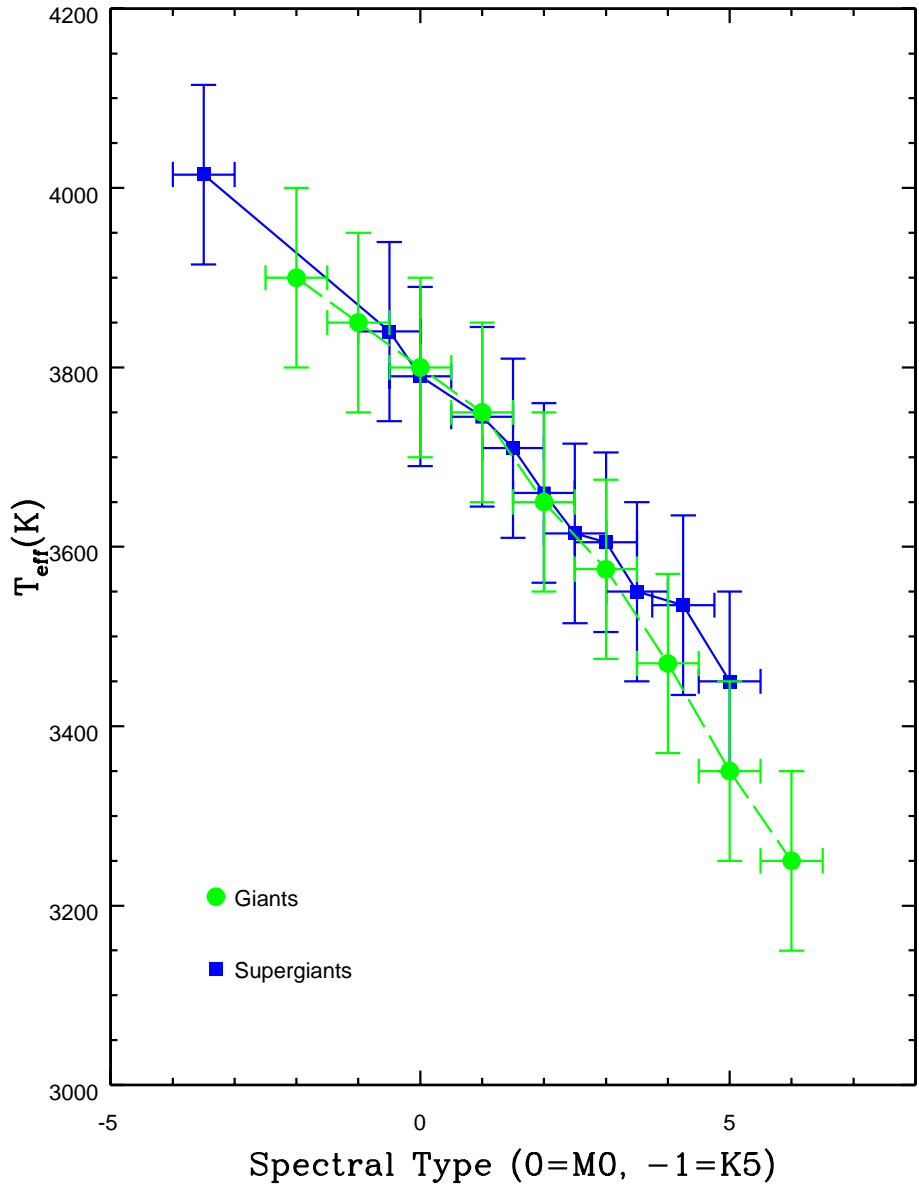}
\figcaption{Effective temperatures as a function of spectral type for both
giants and supergiants.  
The supergiant scale is taken from Levesque et al. (2005),
while the giant scale is defined by effective temperatures from the studies of
Smith \& Lambert (1985; 1986; 1990) coupled to the spectral types from 
Yamashita (1971).  There is a tight relation between T$_{\rm eff}$ and spectral
type, with the giant and supergiant scales not 
so different when defined in terms
of spectral type.  
The error bars illustrate uncertainties of $\pm$100K in T$_{\rm eff}$
and $\pm$0.5 in spectral type.
\label{fig3}}\end{figure}

\begin{figure}
\epsscale{0.79}
\plotone{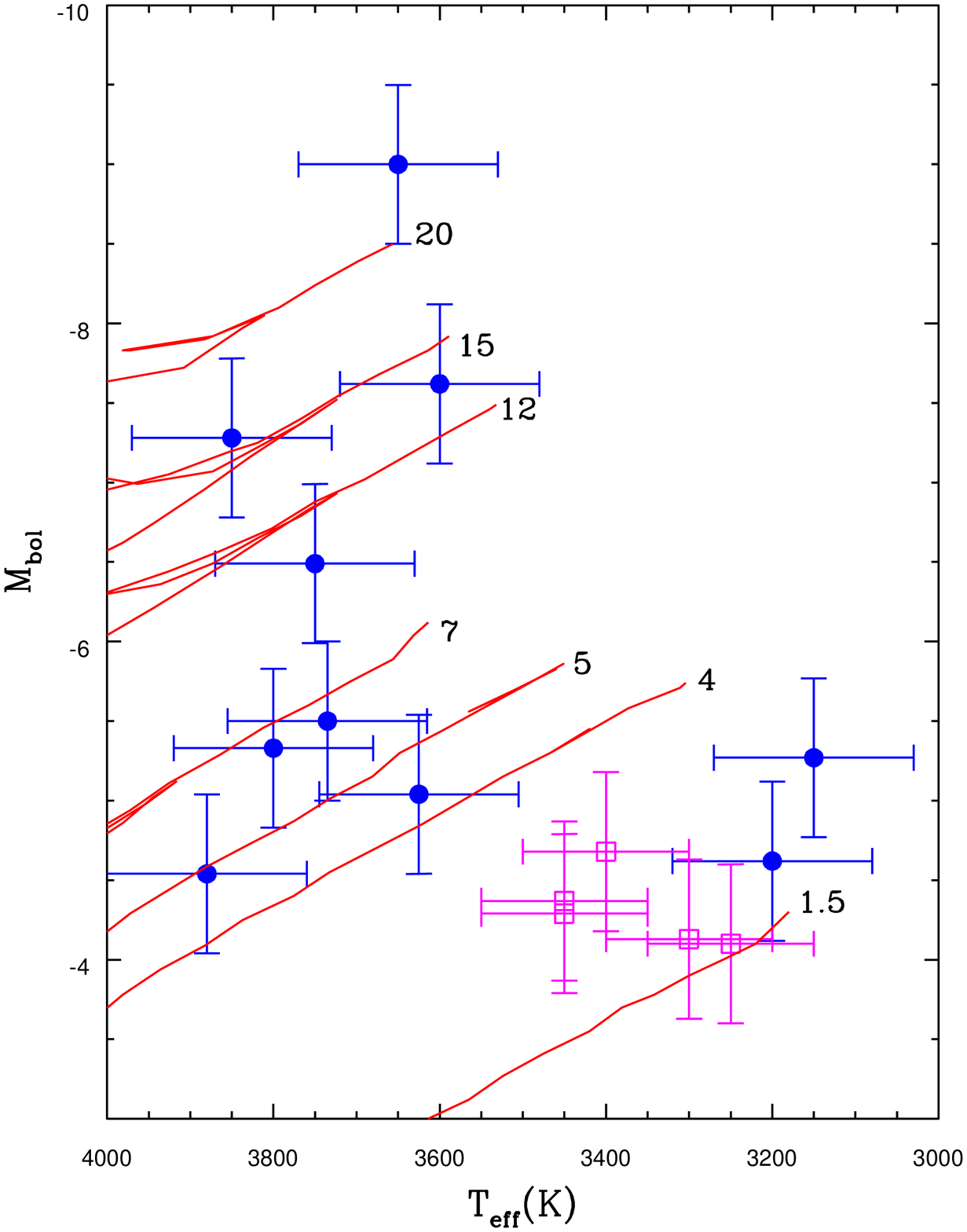}
\figcaption{Absolute bolometric magnitudes and 
effective temperatures for the Galactic
Center stars (blue filled circles) and the 
brighter disk ``standard'' M stars (magenta
open squares).  
Model stellar evolutionary tracks of representative masses are plotted
as the solid red curves, with the model 
stellar mass indicated by each track.  The stellar
model tracks are from Salasnich et al. (2000).    
\label{fig4}}\end{figure}

\begin{figure}
\epsscale{0.79}
\plotone{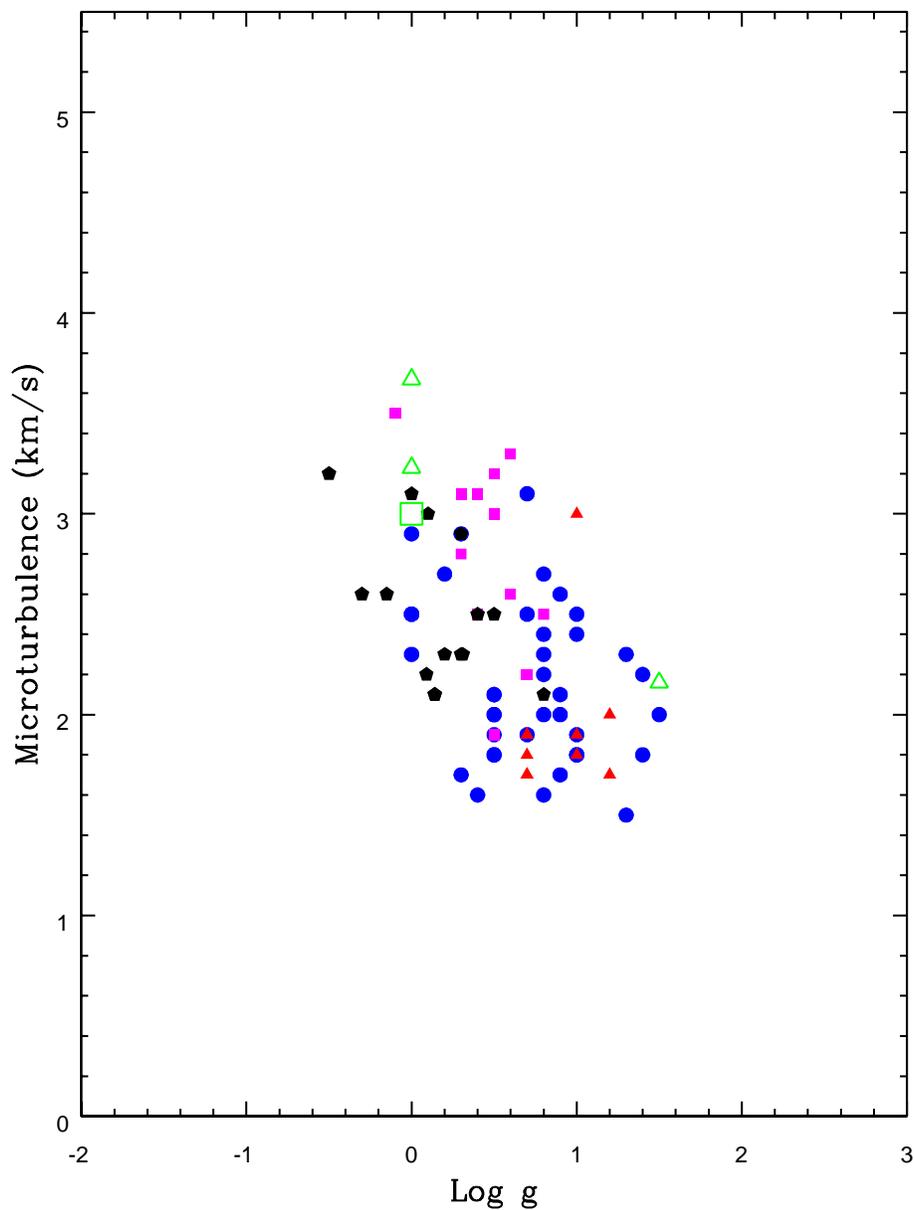}
\figcaption{Microturbulent velocities and surface gravities for
a number of spectroscopic analyses of red giants and supergiants. The results from this
study for both Galactic Center and disk stars are 
represented by black pentagons. The other studies include:
Smith \& Lambert (1986, 1990 - blue solid circles); 
Carr et al. (2000 - green open triangles); 
Smith et al. (2000 - magenta filled squares); 
Smith et al. (2002 - red filled triangles);
Lambert et al. (1984 -  green open square; Alpha Ori).
\label{fig5}}\end{figure}

\clearpage
\begin{figure}
\epsscale{0.59}
\plotone{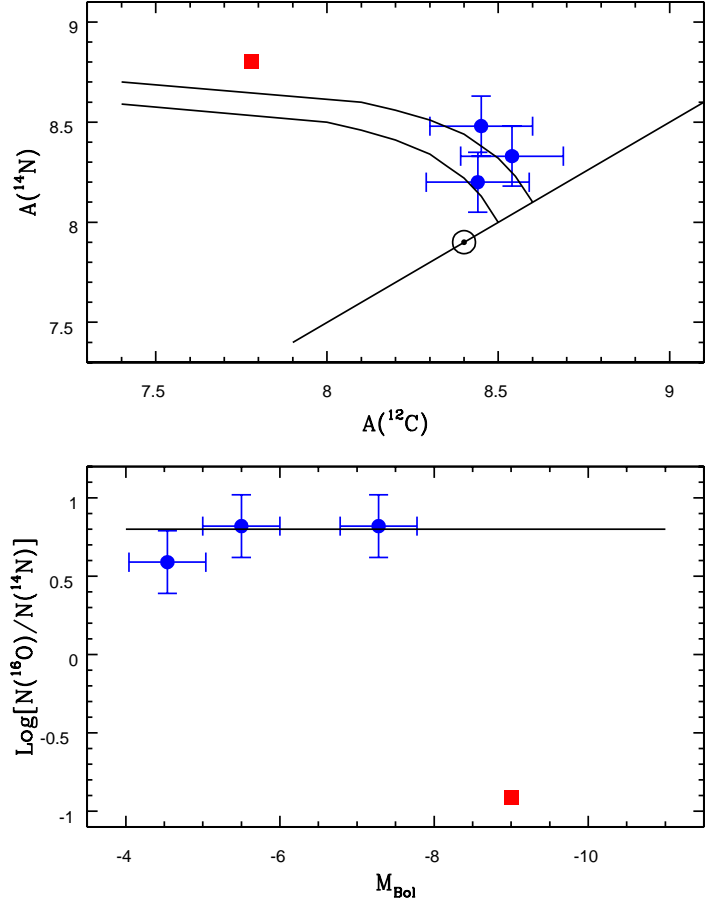}
\figcaption{The top panel shows the nitrogen (A($^{14}$N)) versus
carbon (A($^{12}$C)) abundances for three Galactic Center stars studied
here (filled blue circles), as well as IRS7 (red square; Carr et al. 2000).  
The solar values are shown, along
with a line that defines a solar $^{12}$C/$^{14}$N ratio.  
The solid curves show 
lines of constant $^{12}$C+$^{14}$N, or 
the so-called 'CN-mixing curves', with
initial abundances larger than solar by 0.1 and 0.2 dex, respectively.  
These curves mimic CN-cycle H-burning in which the 
sum of $^{12}$C, $^{14}$N, $^{13}$C,
and $^{15}$N is approximately conserved.  
We do not observe $^{13}$C and
$^{15}$N, but these are minor constituents and 
would change the curves only
slightly.  
The three Galactic Center stars studied here fall within the CN-mixing
curves.  
The very luminous supergiant IRS7 exhibits much more extreme mixing
and the large $^{14}$N abundance may 
require the addition of ON-cycle material
into the atmosphere.  
In the bottom panel $^{16}$O/$^{14}$N ratios versus 
M$_{\rm bol}$ for Galactic Center stars are shown.  
The horizontal line denotes
the solar ratio as a reference point.  
The three evolved stars analyzed here show no
evidence of significant $^{16}$O depletion (from ON-cycle mixing), 
while the
more luminous Galactic Center supergiant IRS7 does show strong evidence for
such mixing.
\label{fig6}}\end{figure}

\clearpage
\begin{figure}
\epsscale{0.45}
\plotone{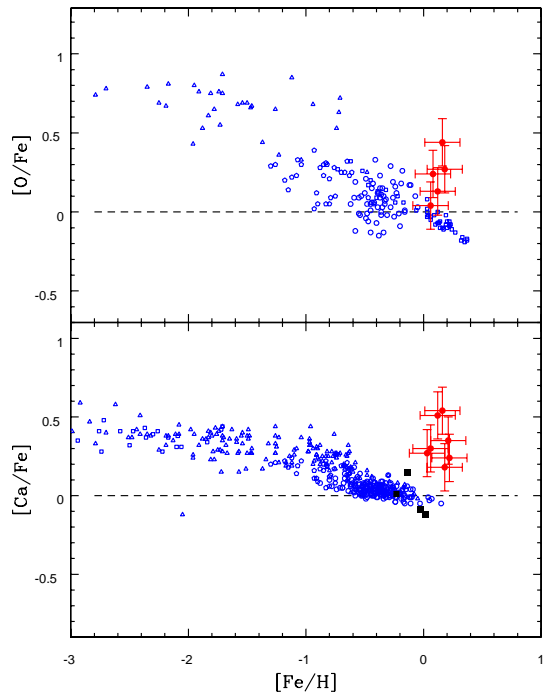}
\figcaption{
The top panel shows the  behavior of [O/Fe] versus [Fe/H] in
different stellar populations.  
The small open symbols (all in blue) are studies of
Galactic disk and halo field stars from Bensby et al. (2004 - squares), 
Reddy et al. (2003 - circles), Nissen \& Schuster (1997 - pentagons), 
and Fulbright \& Johnson (2003 - triangles).  
The decrease in [O/Fe] as [Fe/H] increases is interpreted as
increasing contributions from SN Ia to Fe (but no significant 
O-production from SN Ia)
as time increases, causing [O/Fe] to decline.  
The Galactic Center stars, plotted as
the large filled red circles, are displaced
significantly to larger values of [O/Fe] relative to 
disk stars at the same [Fe/H]; this
suggests a larger relative contribution from SN II chemical 
enrichment in the Galactic
Center population.   
The bottom panel plots the run of [Ca/Fe] values as a function of
[Fe/H].  
The trend of [Ca/Fe] is similar in shape 
to that of [O/Fe] in that [Ca/Fe] is larger
in the more metal-poor stars.  
As calcium, like oxygen, is primarily a product of SN II,
its behavior should track, to some degree, that of oxygen as is observed.  
Again, the
Galactic Center stars stand out from disk stars in having enhanced 
values of [Ca/Fe].
Both Ca and O abundances point to the Galactic Center stars 
as being enriched in SN II chemical yields.  
The filled black squares in this figure are the 4 bright disk
M-giants analyzed for Ca and Fe in the same way as 
for the Galactic Center stars. 
These disk stars follow the same trend in [Ca/Fe] 
versus [Fe/H] as the other disk
samples.  
This result strengthens the conclusion that the Galactic Center 
stars contain elevated values of [Ca/Fe].  
\label{fig7}}\end{figure}

\clearpage

\begin{figure}
\epsscale{0.73}
\plotone{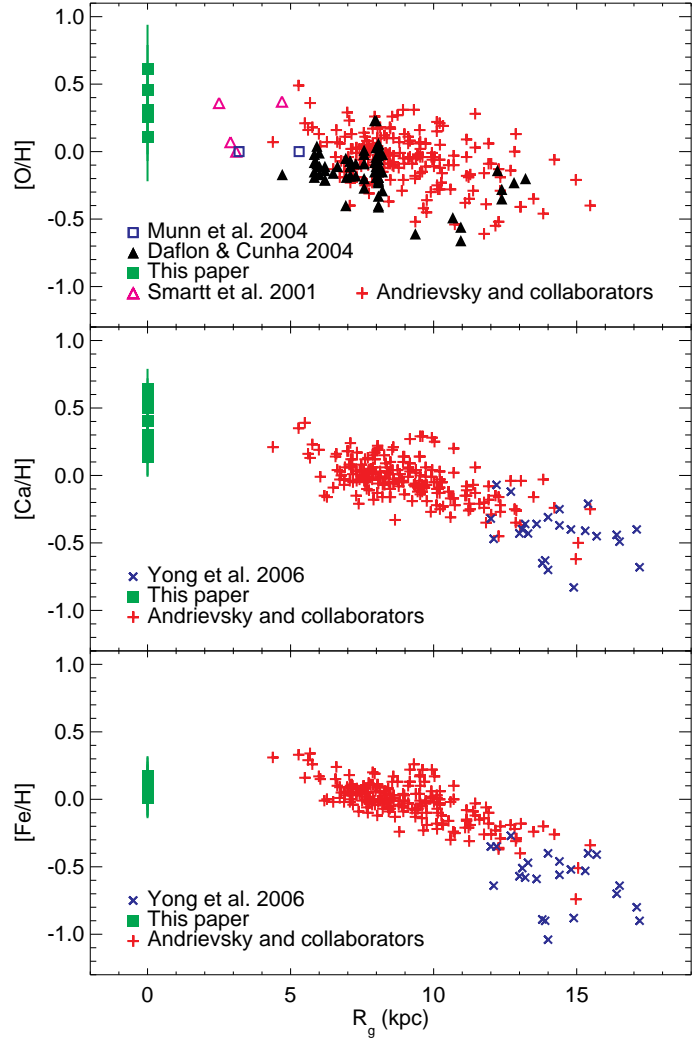}
\figcaption{Abundance versus galactocentric distance,
R$_{g}$, for the elements O ({\it top}), Ca ({\it middle}), 
and Fe ({\it bottom}) for disk stars and Galactic Center stars.  
The disk studies include OB stars for [O/H] (Smartt et al. 2001; 
Daflon \& Cunha 2004; Munn et al. 2004), as well as Cepheids
from Andrievsky and collaborators (Andrievsky et al. 2002a, 2002b, 2002c, 2004;
Luck et al. 2003, 2006; Kovtyukh et al. 2005).
The disk stars for [Ca/H], and [Fe/H] include Cepheids from Andrievsky and collaborators
and from Yong et al. (2006). 
The disk values for [O/H], [Ca/H], and [Fe/H] show increasing abundances with decreasing $R_g$.
The Galactic Center abundances fall near, or below, a smooth extrapolation of the disk trends.
\label{fig8}}\end{figure}

\begin{figure}
\epsscale{1.00}
\plotone{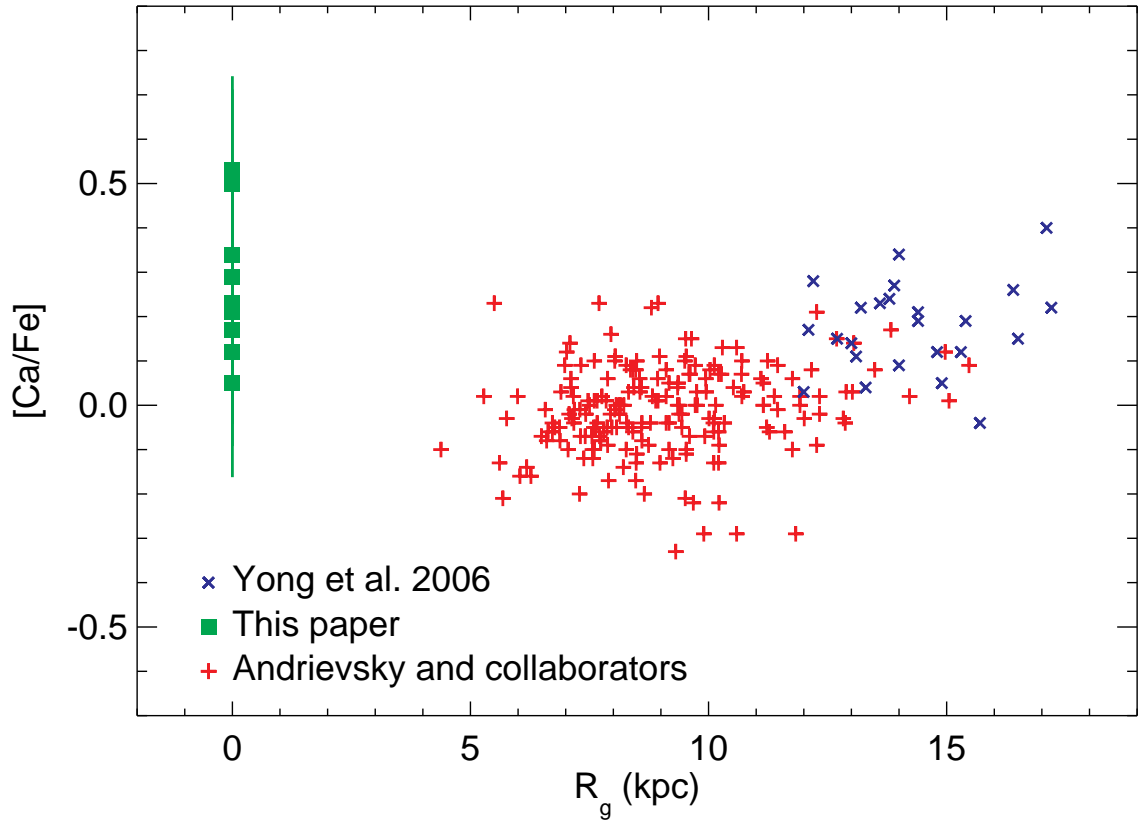}
\figcaption{Calcium to iron abundance ratios (as [Ca/Fe]) for the Cepheid
samples (the same as Figure 8) and the Galactic Center.
Values of [Ca/Fe] in the disk stars decrease as R$_{g}$ decreases, 
whereas the Galactic Center stars have significantly larger values of [Ca/Fe].
It would not appear that the 
Galactic Center Ca and Fe abundances are represented by
smooth extrapolations of the disk trends outside 2.5 kpc.
\label{fig9}}\end{figure}

\end{document}